\documentclass[prl,twocolumn,showpacs,preprintnumbers,amsmath,amssymb,article,superscriptaddress]{revtex4-1}
\pdfoutput=1
\usepackage{graphicx}
\usepackage{float}
\usepackage{verbatim}
\usepackage{dcolumn}
\usepackage{bm}
\usepackage{dsfont}
\usepackage{color}
\usepackage{siunitx}
\usepackage{todonotes}% use \usepackage[disable]{todonotes} to disable notes
\usepackage{physics}

\begin{document}

% MACROS
%%%%%%%%%%%%%%%%%%%%%%%%%%%%%%%%%%%%%%%%%%%%%%%%%%%%%%%%%%%%%%%%%%%%%

\newcommand{\beq}{\begin{equation}}
\newcommand{\eeq}{\end{equation}}
\newcommand{\beqa}{\begin{eqnarray}}
\newcommand{\eeqa}{\end{eqnarray}}
\newcommand{\lf}{\hfil \break \break}
\newcommand{\ahat}{\hat{a}}
\newcommand{\adag}{\hat{a}^{\dagger}}
\newcommand{\adagg}{\hat{a}_g^{\dagger}}
\newcommand{\bhat}{\hat{b}}
\newcommand{\bdag}{\hat{b}^{\dagger}}
\newcommand{\bdagg}{\hat{b}_g^{\dagger}}
\newcommand{\chat}{\hat{c}}
\newcommand{\cdag}{\hat{c}^{\dagger}}
\newcommand{\dhat}{\hat{d}}
\newcommand{\nhat}{\hat{n}}
\newcommand{\ndag}{\hat{n}^{\dagger}}
\newcommand{\den}{\hat{\rho}}
\newcommand{\phihat}{\hat{\phi}}
\newcommand{\Ahat}{\hat{A}}
\newcommand{\Adag}{\hat{A}^{\dagger}}
\newcommand{\Bhat}{\hat{B}}
\newcommand{\Bdag}{\hat{B}^{\dagger}}
\newcommand{\Chat}{\hat{C}}
\newcommand{\Dhat}{\hat{D}}
\newcommand{\Ehat}{\hat{E}}
\newcommand{\Lhat}{\hat{L}}
\newcommand{\Nhat}{\hat{N}}
\newcommand{\Ohat}{\hat{O}}
\newcommand{\Odag}{\hat{O}^{\dagger}}
\newcommand{\Shat}{\hat{S}}
\newcommand{\Uhat}{\hat{U}}
\newcommand{\Udag}{\hat{U}^{\dagger}}
\newcommand{\Xhat}{\hat{X}}
\newcommand{\Zhat}{\hat{Z}}
\newcommand{\Xdag}{\hat{X}^{\dagger}}
\newcommand{\Ydag}{\hat{Y}^{\dagger}}
\newcommand{\Zdag}{\hat{Z}^{\dagger}}
\newcommand{\Ham}{\hat{H}}
\newcommand{\bis}{{\prime \prime}}
\newcommand{\tris}{{\prime \prime \prime}}
\newcommand{\bracket}[3]{\mbox{$\langle#1|#2|#3\rangle$}}
\newcommand{\mat}[1]{\overline{\overline{#1}}}
\newcommand{\dotp}{\mbox{\boldmath $\cdot$}}
\newcommand{\tp}{\otimes}
\newcommand{\hak}[1]{\left[ #1 \right]}
\newcommand{\vin}[1]{\langle #1 \rangle}
\newcommand{\tes}[1]{\left( #1 \right)}
\newcommand{\up}{\ket{\uparrow}}
\newcommand{\down}{\ket{\downarrow}}
\newcommand{\upx}{\ket{x_+}}
\newcommand{\downx}{\ket{x_-}}
\newcommand{\nav}{\langle \hat{n} \rangle}

\hyphenation{Teich}

\title{The crux of using the cascaded emission of a 3--level quantum ladder system to generate indistinguishable photons} %The matter with/

\author{Eva~Sch\"oll}
\thanks{These two authors contributed equally}
\author{Lucas~Schweickert}
\thanks{These two authors contributed equally}
\affiliation{Department of Applied Physics, Royal Institute of Technology, Albanova University Centre, Roslagstullsbacken 21, 106 91 Stockholm, Sweden}
\author{Lukas~Hanschke} %LET ME KNOW THE MUNICH PEOPLE INVOLVED AND ORDER
\affiliation{Walter Schottky Institut and Department of Electrical and Computer Engineering, Technische Universit\"at M\"unchen, 85748, Garching, Germany}
\affiliation{Munich Center of Quantum Science and Technology (MCQST), Schellingstr. 4, 80799 Munich, Germany}
\author{Katharina~D.~Zeuner}
\affiliation{Department of Applied Physics, Royal Institute of Technology, Albanova University Centre, Roslagstullsbacken 21, 106 91 Stockholm, Sweden}%
\author{Friedrich~Sbresny}
\affiliation{Walter Schottky Institut and Department of Electrical and Computer Engineering, Technische Universit\"at M\"unchen, 85748, Garching, Germany}
\affiliation{Munich Center of Quantum Science and Technology (MCQST), Schellingstr. 4, 80799 Munich, Germany}
\author{Thomas~Lettner}
\affiliation{Department of Applied Physics, Royal Institute of Technology, Albanova University Centre, Roslagstullsbacken 21, 106 91 Stockholm, Sweden}%
\author{Rahul~Trivedi}
\affiliation{Ginzton Laboratory, Stanford University, Stanford, CA, USA}
\author{Marcus~Reindl}
\author{Saimon~Filipe~Covre~da~Silva} 
\affiliation{Institute of Semiconductor and Solid State Physics, Johannes Kepler University Linz, 4040, Austria}
\author{Rinaldo~Trotta}
\affiliation{Dipartimento di Fisica, Sapienza Universit\`a di Roma, Piazzale A. Moro 1, I-00185 Roma, Italy}
\author{Jonathan~J.~Finley}
\affiliation{Walter Schottky Institut and Physik Department, Technische Universit\"at M\"unchen, 85748, Garching, Germany}
\affiliation{Munich Center of Quantum Science and Technology (MCQST), Schellingstr. 4, 80799 Munich, Germany}
\author{Jelena~Vu\v{c}kovi\'{c}}
\affiliation{Ginzton Laboratory, Stanford University, Stanford, CA, USA}
\author{Kai~M\"uller} %ALSO CORRESPONDING?
\affiliation{Walter Schottky Institut and Department of Electrical and Computer Engineering, Technische Universit\"at M\"unchen, 85748, Garching, Germany}
\affiliation{Munich Center of Quantum Science and Technology (MCQST), Schellingstr. 4, 80799 Munich, Germany}
\author{Armando~Rastelli}
\affiliation{Institute of Semiconductor and Solid State Physics, Johannes Kepler University Linz, 4040, Austria}
\author{Val~Zwiller}
\author{Klaus~D.~J\"ons}
\email{corresponding author: klausj@kth.se} %WHO WANTS TO BE CORRESPONDING AUTHOR? 
\affiliation{Department of Applied Physics, Royal Institute of Technology, Albanova University Centre, Roslagstullsbacken 21, 106 91 Stockholm, Sweden}%

\date{\today}

\begin{abstract}
We investigate the degree of indistinguishability of cascaded photons emitted from a 3--level quantum ladder system; in our case the biexciton--exciton cascade of semiconductor quantum dots. For the 3--level quantum ladder system we theoretically demonstrate that the indistinguishability is inherently limited for both emitted photons and determined by the ratio of the lifetimes of the excited and intermediate states. We experimentally confirm this finding by comparing the quantum interference visibility of non--cascaded emission and cascaded emission from the same semiconductor quantum dot. Quantum optical simulations produce very good agreement with the measurements and allow to explore a large parameter space. Based on our model, we propose photonic structures to optimize the lifetime ratio and overcome the limited indistinguishability of cascaded photon emission from a 3--level quantum ladder system. 

\end{abstract}

\maketitle

Indistinguishable photons are one of the most essential resources in photonic quantum technologies since they mediate photon--photon interactions by the Hong--Ou--Mandel effect~\cite{Hong.Ou.ea:1987} needed for quantum information processing~\cite{Cirac.Zoller.ea:1997}, quantum sensing~\cite{Giovannetti.Lloyd.ea:2004}, and quantum networks~\cite{Kimble:2008}. Although probabilistic generation of indistinguishable photon pairs by parametric down--conversion has been used as the work horse in quantum optics experiments and in proof--of--principle applications, there is a strong need for the on--demand generation of near--unity indistinguishable photons. This need has led to a whole research field investigating novel solid--state quantum emitters~\cite{Aharonovich.Englund.ea:2016}, optimizing all relevant parameters to fabricate the ideal quantum light source. One promising quantum light source to reach these goals are epitaxially grown semiconductor quantum dots, emitting on--demand~\cite{He.He.ea:2013} near unity indistinguishable single photons~\cite{Somaschi.Giesz.ea:2016,Ding.He.ea:2016}. Quantum dots have recently been used to perform photonic quantum simulations~\cite{Loredo.Broome.ea:2017,He.Ding.ea:2017,Wang.Qin.ea:2019} as well as photonic quantum sensing~\cite{Bennett.Lee.ea:2016,Muller.Vural.ea:2017}. In addition, quantum dots are the only quantum emitter able to generate on--demand polarization entangled photon pairs~\cite{Muller.Bounouar.ea:2014}, using the biexciton--exciton cascade~\cite{Benson.Santori.ea:2000}. This puts quantum dots on the map as ideal quantum light sources to realize quantum relays~\cite{Huwer.Stevenson.ea:2017,Basset2019,Zopf2019} and quantum repeaters based on the Shapiro Lloyd scheme~\cite{Lloyd.Shahriar.ea:2001}. However, these applications require the simultaneous generation of near--unity indistinguishable and maximally entangled photon pairs. Despite, large research efforts~\cite{Muller.Bounouar.ea:2014, Stevenson.Salter.ea:2012, Reindl.Joens.ea:2017, Huber.Reindl.ea:2017} achieving highly indistinguishable photons from the biexciton--exciton cascade has proven to be elusive even under optimized excitation conditions~\cite{Huber_2015}. Here, we show that this stems from the intrinsic properties of the quantum 3--level ladder system, inherent to quantum dots, which reduces the maximally achievable indistinguishability of the emitted photons. To this end, we first demonstrate analytically that the indistinguishability is identical for the emission from either of the cascaded transitions, albeit limited. This finding is then tested experimentally by measuring the photon indistinguishability for emission from a quantum 2--level system and a quantum 3--level ladder system using the same quantum emitter. Here, we evaluate data from four different quantum dots and extract the relevant parameters. Finally, we perform quantum--optical simulations exploring a wide parameter space. The results are in good agreement with our measurements. To overcome the limitations of 3--level quantum ladder systems we propose nano--engineering the lifetimes of the involved excited states.

%Theory
%\todo[inline]{BEGIN THEORY}
To gain theoretical insight into the fundamental limits of the trace purity of photons obtained from the cascaded two-photon emission, we consider a 3--level quantum ladder system with ground state $\ket{\text{G}}$, intermediate state $\ket{\text{X}}$ and excited state $\ket{\text{XX}}$. Using a short laser pulse, the system is initialized to the excited state $\ket{\text{XX}}$ at $t=0$ and allowed to decay from $\ket{\text{XX}} \to \ket{\text{X}}\to \ket{\text{G}}$ via the cascaded emission of two photons. The emitted two-photon state can be easily computed using a scattering matrix formalism \cite{Fischer2018c, trivedi2018few} as:
\begin{subequations}
\begin{align}\label{eq:pure_state}
    \ket{\psi} = \int_{t=0}^\infty \int_{t'=t}^\infty \dd{t} \dd{t'} f(t, t') b_{\text{XX}\to \text{X}}^\dagger(t)b_{\text{X}\to \text{G}}^\dagger(t') \ket{\text{vac}; \text{G}},
\end{align}
where $b^\dagger_{\text{XX}\to \text{X}}(t)$ and $b^\dagger_{\text{X}\to \text{G}}(t)$ are the time-domain creation %annihilation
operators describing the photonic modes that the transitions $\ket{\text{XX}}\to \ket{\text{X}}$ and $\ket{\text{X}}\to \ket{\text{G}}$ couple to, and $f(t, t')$ is the two-photon wavefunction of the emitted photon given by
%where $b_{e\to i}(t)$ and $b_{i\to g}(t)$ are the time-domain annihilation operators 
\begin{align}
    &f(t, t') =\nonumber\\
    &\sqrt{\gamma_{\text{X}} \gamma_{\text{XX}}} e^{-\textrm{i}\omega_{\text{XX}\to \text{X}} t}e^{-\textrm{i}\omega_{\text{X}\to \text{G}} t'} e^{-\gamma_{\text{X}} (t' - t)/2}e^{-\gamma_{\text{XX}} t/2},
\end{align}
\end{subequations}
where $\omega_{\text{XX}\to \text{X}}, \omega_{\text{X}\to \text{G}}$ are the frequencies of the transitions $\ket{\text{XX}}\to \ket{\text{X}}, \ket{\text{X}} \to \ket{\text{G}}$ and $\gamma_{\text{XX}}, \gamma_{\text{X}}$ are the decay rates of the states $\ket{\text{XX}}, \ket{\text{X}}$. The single photon emitted from the transition $\ket{\text{XX}}\to \ket{\text{X}}$ can then be described by a mixed state obtained by tracing the pure two-photon state ($\ket{\psi}\bra{\psi}$) over the modes described by operator $b_{\text{X}\to \text{G}}(t)$. The density matrix of this state is calculated from Eq.~\ref{eq:pure_state}:
\begin{subequations}\label{eq:density_matrix}
\begin{align}
 &\rho = \text{Tr}_{b_{\text{X}\to \text{G}}}\big[ \ket{\psi}\bra{\psi}\big] \nonumber\\
 &= \int_0^\infty \int_0^\infty \dd{t} \dd{t'} \rho(t, t') b_{\text{XX}\to \text{X}}^\dagger(t)\ket{\text{vac}}\bra{\text{vac}}b_{\text{XX}\to \text{X}}(t'),
\end{align}
where
\begin{align}
    \rho(t, t') = \gamma_{\text{XX}} e^{\textrm{i}\omega_{\text{XX}\to \text{X}}(t' - t)} e^{-\gamma_{\text{XX}}(t + t') / 2}e^{-\gamma_{\text{X}} |t - t'|/2}
\end{align}
\end{subequations}
It can be noted from Eq.~\ref{eq:density_matrix} that the state of the single photon being emitted via the transition $\ket{\text{XX}}\to \ket{\text{X}}$, while generally being a non-separable state, becomes separable if the decay rate of the intermediate state $\ket{\text{X}}$ vanishes. The non-separability of the emitted photon immediately limits the indistiguishability of photons emitted from such a cascaded 3-level system, since for the emitted photons to be identical they must be describable as pure states and thus separable. The separability of this single-photon state can be quantified by its trace purity $\mathbb{P}$ which can be analytically evaluated using Eq.~\ref{eq:density_matrix}~\cite{Simon.Poizat:2005}:
\begin{align}
    \mathbb{P} = \text{Tr}_{b_{\text{XX}\to \text{X}}}\big[\rho^2\big] = \frac{\gamma_{\text{XX}}}{\gamma_{\text{XX}} + \gamma_\text{X}}
\end{align}
For quantum dots where typically the emission of the biexciton has twice the rate of emission from the exciton $\gamma_{\text{XX}} = 2\gamma_\text{X}$, the maximum achievable trace purity is limited to $\sim 0.66$.

While the trace purity described above theoretically describes the indistinguishability of the emitted single-photon, it is not possible to measure it directly in experiments. A more experimentally accessible quantity is the interference visibility parameter $v$ extracted from two-photon interference experiments \cite{Fischer2018, trivedi2019generation}. However, for systems with negligible emission with photon numbers $>1$, it has been recently shown that $\mathbb{P}$ and $v$ are identical~\cite{Fischer2018} and consequently $v$ can be taken to be a measure of the indistinguishability of the emitted photons.

\begin{figure}
    \centering
    \includegraphics[width=\linewidth]{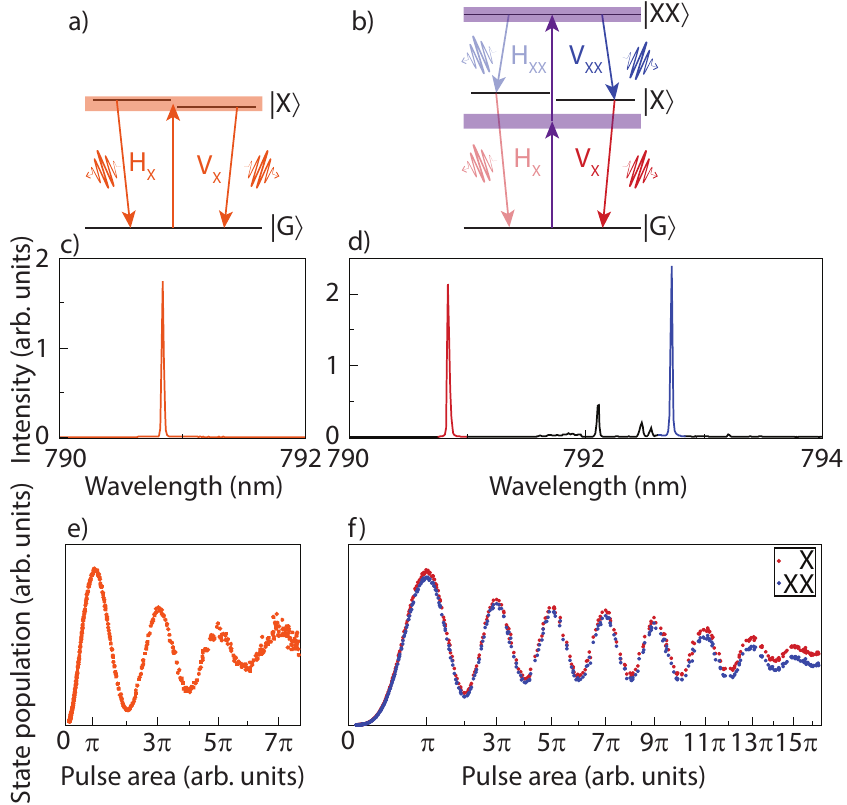}
    \caption{Characterization of the quantum dot under s--shell resonant excitation (a, c, e) and two--photon excitation (b, d, f). a) s--shell resonant excitation scheme. The exciton level ($\ket{\text{X}}$) is directly addressed by the excitation laser. Under recombination into the ground state ($\ket{\text{G}}$) a resonance fluorescence (RF) photon is emitted.  b) Two--photon excitation (TPE) scheme. The biexciton state ($\ket{\text{XX}}$) is resonantly driven via a two--photon process, which recombines via the exciton state into the ground state, resulting in a biexciton (blue) --  exciton (red) cascade. The H polarized photons are shaded to illustrate, that one polarization component is suppressed by the cross--polarization setup. c) Resonance fluorescence spectrum of the exciton. d) Two--photon excitation spectrum with the exciton (red) at lower wavelength compared to the biexciton (blue).  Excitation laser power--dependent Rabi oscillation up to a pulse area of e) $7\,\pi$ under s--shell resonant excitation and f) $16\,\pi$ for two--photon excitation.}
    \label{fig:expRabi}
\end{figure}

In the following we compare a quantum 2--level system and a 3--level quantum ladder system, by measuring the second--order coherence and two--photon interference of four different semiconductor quantum dots. 
Our 2--level quantum system is an exciton state in a semiconductor quantum dot, which is directly addressed in a pure s--shell resonant excitation scheme, depicted in Fig.\,\ref{fig:expRabi}\,a. When the system recombines back to the ground state, a resonance fluorescence photon is emitted.
The experimentally investigated 3--level quantum ladder system is the biexciton--exciton cascade of semiconductor quantum dots. 
In this system, the biexciton state is resonantly addressed in a two--photon process (two--photon excitation) and recombines via the exciton state into the ground state resulting in the characteristic biexciton--exciton cascade, shown in Fig.\,\ref{fig:expRabi}\,b. This cascade is crucial to generate entangled photon pairs and has been extensively studied in literature~\cite{Huber.Predojevic.ea:2013,Ward.Dean.ea:2014,Orieux.Versteegh.ea:2017}.

We use GaAs/AlGaAs quantum dots grown via the droplet--etching method~\cite{Huo.Rastelli.ea:2013}. A detailed description of the sample structure can be found in~\cite{Huber.Reindl.ea:2017}. Remarkably these quantum dots currently hold the record for highest degree of entanglement when generating polarization entangled photon pairs~\cite{Huber.Reindl.ea:2018}, lowest multi--photon emission probability~\cite{Schweickert.Joens.ea:2018} and are among the brightest entangled photon sources to date~\cite{Liu.Su.ea:2019,Wang.Hu.ea:2019}.

All experiments are performed in a confocal micro--photoluminescence spectroscopy setup~\cite{Schoell.Hanschke.ea:2019}, (see supplementary material~\cite{Suppl:2020} for more details and a schematic). The quantum dot sample is mounted in a closed--cycle low vibration cryostat and cooled down to \SI{5}{\kelvin}. The quantum dot of interest is excited with \SI{5}{\pico\second} long laser pulses with a repetition rate of \SI{80}{\mega\hertz}. For filtering of the signal, a polarization suppression setup is used similar to Ref.~\cite{Kuhlmann.Houel.ea:2013}. The signal can either be detected by the CCD of the spectrometer for photoluminescence measurements or further filtered in a home--built transmission spectrometer for correlation spectroscopy. The second--order autocorrelation is measured in a Hanbury Brown and Twiss type experiment whereas the two--photon interference visibility is measured using a Hong--Ou--Mandel setup.
%To measure the degree of second--order coherence, the second--order autocorrelation is measured in a Hanbury Brown and Twiss type experiment, where the signal is split up in a 50:50 beam splitter where the outputs are detected by two superconducting single photon detectors and correlated using a time tagging module and the Extensible Timetag Analyzer (ETA)~\cite{ETA}. The two--photon interference visibility is measured using a Hong--Ou--Mandel type experiment. For this, the signal is sent into an unbalanced Mach--Zehnder--interferometer with a fixed delay of $\tau_0=2\,\si{\ns}$, where the outputs are detected and correlated as in a Hanbury Brown and Twiss type measurement. To achieve temporal overlap on the second beam splitter, the excitation pulses are split into double pulses by sending the laser through another unbalanced Mach--Zehnder--interferometer with a variable delay stage set to the same delay as for the single photon interferometer.

%\begin{figure}
%    \centering
%    \includegraphics[width=\linewidth]{XXX}
%    \caption{}
%    \label{fig:theory}
%\end{figure}

\begin{figure}
    \centering
    \includegraphics[width=\linewidth]{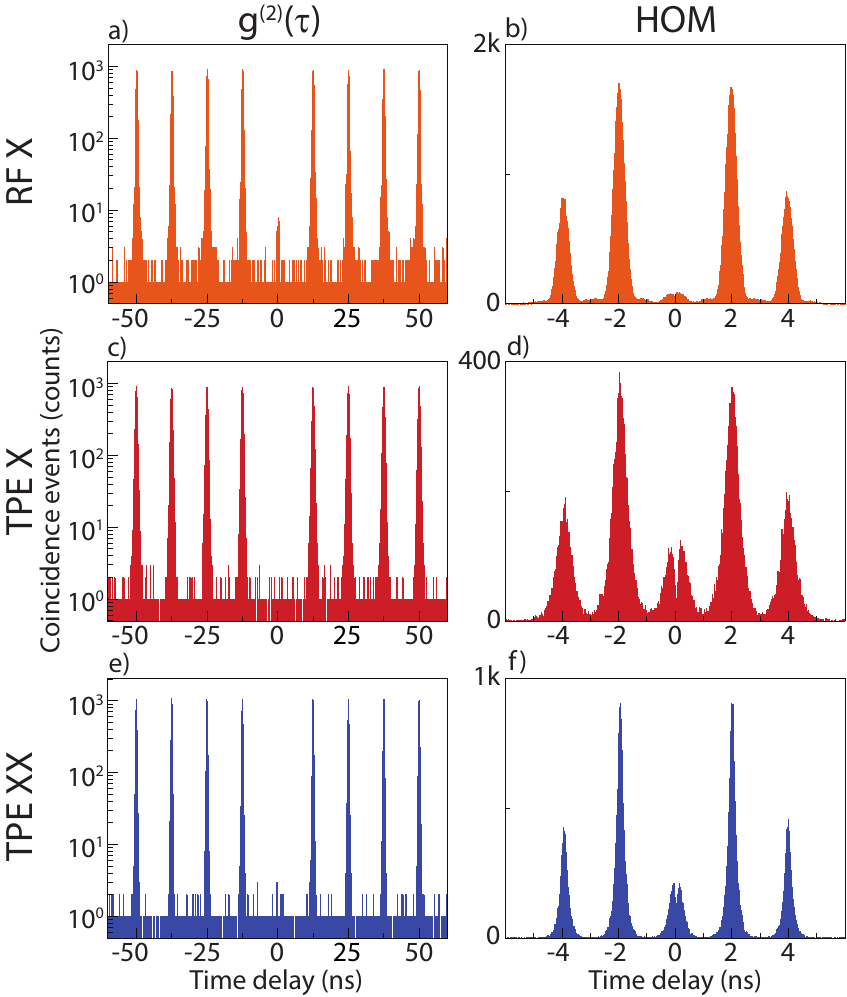}
    \caption{Second--order intensity correlation histogram of the a) exciton (X) under s--shell resonant excitation (RF) with g$^{(2)}_{\text{RF,X}}(0)=(8.16\pm0.55)\times 10^{-3}$, two--photon excitation (TPE) of the c) exciton with g$^{(2)}_{\text{TPE,X}}(0)=(9.13\pm1.61)\times 10^{-4}$ and e) the biexciton (XX) with g$^{(2)}_{\text{TPE,XX}}(0)=(1.79\pm0.30)\times10^{-3}$. Hong--Ou--Mandel (HOM) histograms of the b) exciton emission under resonant excitation $v_{\text{RF,X}}=92.3\pm 0.1\,\%$, two--photon excitation of the d) exciton with $v_{\text{TPE,X}}=56.7\pm 0.6\,\%$ and f) the biexciton $v_{\text{TPE,XX}}=60.0\pm 0.5\,\%$.}
    \label{fig:expHOM}
\end{figure}

%In the following we compare a quantum mechanical 2--level system and a 3--level ladder system, by measuring the second--order coherence and two--photon interference of four different quantum dots, whereas only the data of QD 1 is shown here. In the 2--level system, the excited exciton state is directly addressed in a pure s--shell resonant excitation scheme, depicted in Fig.~\ref{fig:expRabi}a. When the system recombines back to the ground state, a resonance fluorescence (RF) photon is emitted.In the 3--level quantum ladder system, the higher energetic biexciton state is addressed in a two--photon process (two--photon excitation, TPE) , shown in Fig.~\ref{fig:expRabi}b. Hence, the exciton state gets indirectly populated by the biexciton state, resulting in the characteristic biexciton--exciton cascade.

The spectrum of the exciton under pure s--shell resonant excitation (Fig.\,\ref{fig:expRabi}\,c) consists of a single sharp peak. Power--dependent resonance fluorescence spectroscopy (Fig.\,\ref{fig:expRabi}\,e) reveals clean Rabi oscillation up to $7\,\pi$ proving the coherence of this excitation scheme.
The same measurements for the exciton and biexciton under two--photon excitation are shown in Fig.\,\ref{fig:expRabi}\,d and f. In the spectrum, two lines for the exciton (red) and biexciton (blue) can be seen with a wavelength difference of \SI{1.9}{\nano\meter} (\SI{3.7}{\milli\electronvolt}), which stems from the Coulomb interaction between the two excitons, resulting in an energetically lower (longer wavelength) biexciton photon. In between those two excitonic lines are some other lines stemming from the quantum dot, which are additionally spectrally filtered for the following correlation measurements by the transmission spectrometer. In this excitation scheme, Rabi oscillations up to $16\,\pi$ can be distinguished. The difference to the s--shell resonant excitation stems from the fact that here the laser can be filtered spectrally in addition to cross--polarization, allowing much higher excitation powers.
All further measurements are performed with an excitation power corresponding to a pulse area $\pi$, where the system is maximally inverted.
Figure\,\ref{fig:expHOM} shows correlation measurements for both quantum systems. The second--order autocorrelation function is shown in Fig.\,\ref{fig:expHOM}\,a,c,e in a semi--logarithmic plot. For all measurements, the peak at time delay zero is strongly suppressed proving almost background--free single--photon emission. Exact results are shown in Tab.\,1 of the supplementary material~\cite{Suppl:2020}. For all four quantum dots we investigated, the measurements under two--photon excitation show lower values than for pure s--shell resonant excitation, which we attribute to the suppressed re--excitation processes ~\cite{Schweickert.Joens.ea:2018, Hanschke2018}. 
%Furthermore in the latter scheme, for all quantum dots but QD\,1, the biexciton shows lower coherence than the exciton.
%, yielding g$^{(2)}(0)=(8.16\pm0.55)\times 10^{-3}$ for the exciton under s--shell resonant excitation (Fig.~\ref{fig:expHOM}a), g$^{(2)}(0)=(9.13\pm1.61)\times 10^{-4}$ the exciton under two--photon excitation (Fig.~\ref{fig:expHOM}b), and g$^{(2)}(0)=(1.79\pm0.30)\times10^{-3}$ for the biexciton under two-photon excitation (Fig.~\ref{fig:expHOM}c).
Next, we investigate the indistinguishability by measuring the two--photon interference of two consecutively emitted photons in a Hong--Ou--Mandel type experiment under the same excitation conditions as above. The measurement result is presented in Fig.\,\ref{fig:expHOM}\,b,d,f, where the suppression of the center peak is a measure for the visibility $v$ and characterizes the indistinguishability of two photons.
The experimental methods and results of all quantum dots under both excitation schemes are summarized in Tab.\,1 of the supplementary material~\cite{Suppl:2020}. %In addition, the two--photon interference measurements of QD\,2 are presented in Fig.\,\ref{fig:expHOM}\,g--i.
The raw Hong--Ou--Mandel visibility is above $90\,\%$ for all quantum dots under pure s--shell resonant excitation, whereas it maximally reaches $64\,\%$ under two--photon excitation, consistent with theoretical predictions for a cascaded emission as discussed above.

\begin{figure}
	\centering
	\includegraphics[width=\linewidth]{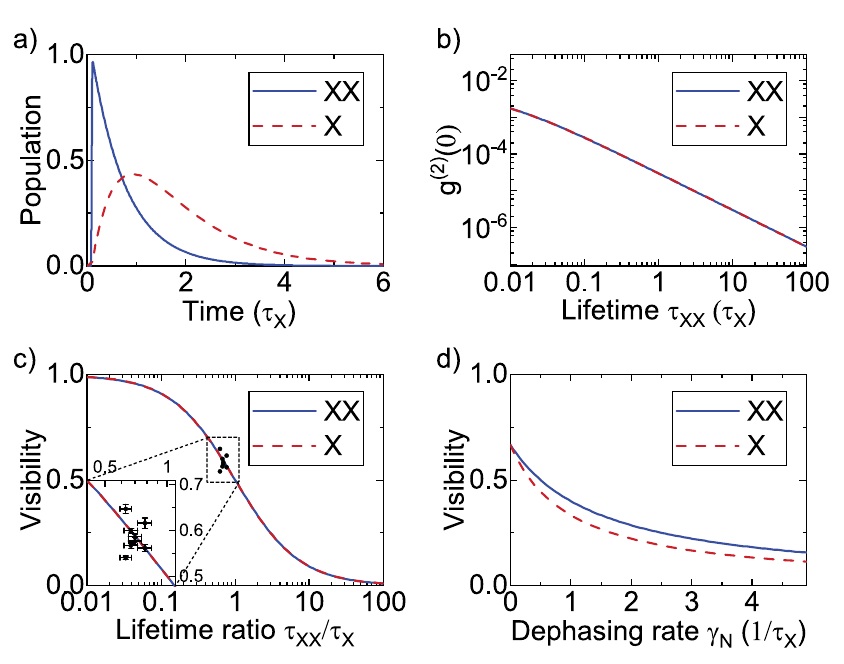}
	\caption{Simulation of the two--photon excitation of a 3--level quantum ladder system with a Gaussian pulse of duration (FHWM) $\tau_{\text{X}}/50$ which is a commonly used pulse length in quantum dot experiments. a) Evolution of the biexciton (blue) and exciton (red) state population excited by a $\pi$--pulse as a function of the exciton lifetime. The exciton population builds up as the biexciton decays. b) Simulated second--order coherence function at zero time delay $g^{(2)}(0)$ for different biexciton lifetimes. With increasing $\tau_{\text{XX}}$ the $g^{(2)}(0)$ value decreases by three orders of magnitude for a fixed exciton lifetime $\tau_{X}$. c) Hong--Ou--Mandel visibility of the two transitions in dependence of the lifetime ratio. The visibility decreases drastically from near unity for increasing $\tau_{\text{XX}}/\tau_{\text{X}}$. Inset: zoom-in showing experimental data close to the simulated curve. d) By adding a phenomenological pure dephasing acting with equal strength on both transitions for a fixed value of $\tau_{\text{XX}}/\tau_{\text{X}}=0.5$, the obtainable Hong--Ou--Mandel visibility of both transitions decreases with a higher impact on the exciton transition visibility.}
	\label{fig:simulations}
\end{figure}

%Simulations
%\todo[inline]{SIMULATION PART}
In order to obtain a deeper insight into how the biexciton--exciton cascade results in a reduced Hong--Ou--Mandel visibility, we performed quantum--optical simulations using the Quantum Toolbox in Python (QuTiP)~\cite{Johansson2012}. We model the cascade as a 3--level quantum ladder system and use a quantum--optical master equation approach described in detail in~\cite{Fischer2016}. These simulations go beyond the analytical theoretical considerations above as they take into account the excitation laser pulse width as well as dephasing. We neglect the fine structure splitting in the simulation since one fine structure channel is suppressed due to the cross--polarized resonance fluorescence setup. The Hamiltonian in the rotating frame at the laser frequency then reads
\begin{equation}
    H(t)= \frac{\left(\mu \cdot E(t)\right)^2}{2E_b} \left(\ket{\text{G}}\bra{\text{XX}}+ \ket{\text{XX}}\bra{\text{G}}\right)
\end{equation}
where $\mu$ is the electric dipole moment, $E(t)$ the electric field and $E_b$ is the binding energy of the biexciton. The crystal ground state $\ket{\text{G}}$ and the biexciton level $\ket{\text{XX}}$ are coupled by the electromagnetic field of a Gaussian laser pulse of length (FWHM) $\tau_{\text{X}}/50$, where $\tau_{\text{X}}$ denotes the exciton lifetime. The pulse area
\begin{equation}
    A(t) = \int_{0}^{t}\dd{t'} \frac{\left(\mu \cdot E(t')\right)^2}{\hbar E_b}
\end{equation}
is set to $\pi$ to achieve a maximum population inversion. The two radiative decays between the three levels $\ket{\text{XX}} \rightarrow \ket{\text{X}}$ and $\ket{\text{X}} \rightarrow \ket{\text{G}}$ are coupled via two collapse operators $c_{\text{XX}}=\sqrt{1/\tau_{\text{XX}}}\,e$ and $c_{\text{X}}=\sqrt{1/\tau_{\text{X}}}\,i$, where $e=\ket{\text{X}}\bra{\text{XX}}$ and $i=\ket{\text{G}}\bra{\text{X}}$ are the lowering operators of the excited and the intermediate state. We disregard any non--radiative decay mechanisms, as they do not contribute to the photon statistics measurements. The population evolution of this system is depicted in Fig.\,\ref{fig:simulations}\,a. During one excitation pulse the biexciton population builds up, reaching nearly unity. It then decays into the intermediate exciton state under the emission of a photon. This leads to a buildup of the exciton population, which decays into the ground state emitting another photon.

Having set up the dynamical quantum system, we turn our interest to how the properties of the emitted photons depend on the ratio of the two transition lifetimes $\tau_{\text{XX}}/\tau_{\text{X}}$. To investigate the single photon character, we calculate the second--order correlation function normalized by the total photon number via the equation 
\begin{equation}
g^{(2)}(0)= \frac{2 \int_{0}^{T}\int_{0}^{T}\dd{t}\dd{\tau}\langle s^{\dagger}(t)s^\dagger(t+\tau)s(t+\tau)s(t) \rangle}{\left( \int_{0}^{T}\dd{t}\langle s^{\dagger}(t)s(t)\rangle \right)^2}
\end{equation}
where $s$ and $s^{\dagger}$ are the lowering and raising operators of the studied transition. The factor 2 is included to also take into account negative values of $\tau$. The value of $g^{(2)}(0)$ is plotted in Fig.\,\ref{fig:simulations}\,b as a function of the lifetime $\tau_{\text{XX}}$ for a constant lifetime $\tau_{\text{X}}$. With increasing $\tau_{\text{XX}}$ the value of $g^{(2)}(0)$ decreases by three orders of magnitude. As $\tau_{\text{X}}$ is fixed, increasing $\tau_{\text{XX}}$ increases the total time that it takes the system to return to the ground state, which reduces re--excitation during the presence of the laser pulse and thus multi--photon emission~\cite{Hanschke2018, Schweickert.Joens.ea:2018}. For the dots studied in this work the lifetime ratio is $0.78 \pm 0.06 > \tau_{\text{XX}}/\tau_{\text{X}} > 0.63 \pm 0.04$ which results in values on the order of $g^{(2)}(0)=10^{-5}$, listed in Tab.\,1 of the supplementary material~\cite{Suppl:2020} for comparison to the measured values.% Tuning the ratio over a wide range could be experimentally realized by embedding the quantum dot into a nanophotonic resonator like photonic crystals~\cite{Lodahl.FlorisvanDriel.ea:2004}, micropillars~\cite{PhysRevLett.103.027401}, bull's eye cavities~\cite{Kolatschek.Hepp.ea:2019}, paraboloids~\cite{Lettner.Zeuner.ea:2020}, and even planar cavities~\cite{Huber.Predojevic.ea:2013} which selectively enhances / reduces the lifetimes $\tau_{\text{XX}}$ and $\tau_{\text{X}}$. We would like to note that asymmetric Purcell enhancement of both transitions is preferred over Purcell suppression to maintain efficient photon extraction from the cavity.

We continue by simulating the indistinguishability of the emitted photons in a Hong--Ou--Mandel (HOM) interferometer as outlined in reference ~\cite{Fischer2016}
\begin{equation}
g^{(2)}_{\text{HOM}}(0)= \frac{1}{2} g^{(2)}(0)+\frac{1}{2}\left( 1-\left| g^{(1)}(0) \right|^2 \right)
\end{equation}
In this normalization, for a non--ideal single--photon source with $g^{(2)}(0)>0$ the value $g^{(2)}_{\text{HOM}}(0)$ would increase as the beam splitter with the two detectors simultaneously represents a Hanbury Brown and Twiss experiment for each excitation pulse.  
As $g^{(2)}(0) \approx 0 $ both in experiment and simulation for the studied parameter space, we focus on the visibility  
\begin{equation}
v= \left| g^{(1)}(0) \right|^2 =  \frac{2\int_{0}^{T}\int_{0}^{T} \dd{t}\dd{\tau} \left| \langle s^{\dagger}(t+\tau) s(t) \rangle \right|^2 }{\left( \int_{0}^{T}\dd{t}\langle s^{\dagger}(t)s(t)\rangle \right)^2}
\end{equation}
The dependence of the visibility parameter with respect to the lifetime ratio is presented in Fig.\,\ref{fig:simulations}\,c. It decreases from near unity for small values of $\tau_{\text{XX}}/\tau_{\text{X}}$ to almost zero for large ratios. The experimental values of approximately $60\,\%$  (Inset Fig.\,\ref{fig:simulations}\,c) match  the simulated curve well. Tuning the ratio over a wide range could be experimentally realized by embedding the quantum dot into a nanophotonic resonator like photonic crystals~\cite{Lodahl.FlorisvanDriel.ea:2004}, micropillars~\cite{PhysRevLett.103.027401}, bull's eye cavities~\cite{Kolatschek.Hepp.ea:2019}, paraboloids~\cite{Lettner.Zeuner.ea:2020}, and even planar cavities~\cite{Huber.Predojevic.ea:2013} which selectively enhances / reduces the lifetimes $\tau_{\text{XX}}$ and $\tau_{\text{X}}$. We would like to note that asymmetric Purcell enhancement of both transitions is preferred over Purcell suppression to maintain efficient photon extraction from the cavity.
The simulation results for the lifetime ratio of our measured quantum dots are given in Tab.\,1 in the supplementary material~\cite{Suppl:2020}.
%For the studied quantum dots, the simulated visibilities given in Tab.\,1 of the supplementary material of approximately 60\% closely matches the experimental results for the corresponding lifetimes. 
As discussed above, the limited visibilities for both transitions result from the entangled nature of the cascaded emission. However, it can also be explained in a simple picture by the finite lifetime of the exciton. The photons emitted by the biexciton are spectrally broadened due to the linewidth of its final state ($\ket{\text{X}}$) resulting from its finite lifetime. On the other hand photons emitted from the exciton state are subject to a timing jitter induced by the cascade.

Finally, we investigate how introducing additional dephasing affects the two--photon interference visibilities of both transitions. To this end, we introduce a phenomenological pure dephasing acting with equal strength on both levels, described by two additional collapse operators of the form $c_{N}^{\text{XX}}= \sqrt{\gamma_N} \ket{\text{XX}}\bra{\text{XX}}$ and $c_{N}^{\text{X}}= \sqrt{\gamma_N} \ket{\text{X}}\bra{\text{X}}$. The visibility of both transitions is presented in Fig.\,\ref{fig:simulations}\,d as a function of the dephasing rate $\gamma_N$ for a fixed ratio of $\tau_{\text{XX}}/\tau_{\text{X}}=0.5$. With increasing dephasing rate, the visibility of both transitions decreases whereby the visibility of the exciton emission (red) is lower than that of the biexciton emission (blue). This stems from the faster biexciton emission rate compared to that of the exciton, so that the biexciton state is less affected by the same dephasing rate. This is also consistent with our experimental measurements discussed above, where typically a higher visibility was measured for the biexciton transition.

%Conclusion
In summary, we investigated the impact of the intermediate state on the indistinguishability of cascaded photons emitted by a 3--level quantum ladder system and experimentally confirmed our findings using the biexciton--exciton cascade in semiconductor quantum dots. The key parameter for reaching near-unity indistinguishability for photons stemming from a cascaded emission is the ratio between the lifetimes of the two excited states. Therefore, asymmetric Purcell enhancement of these states is expected to overcome the limitation on the indistinguishability of cascaded photons emitted by a 3--level quantum ladder system, bringing the indistinguishability to the dephasing-limited values of a resonantly driven two-level system. This would enable the simultaneous generation of near-unity indistinguishable and entangled photon pairs, required for photonic entanglement-based quantum repeater schemes.  

\begin{acknowledgments}
This project has received funding from the European Union's Horizon 2020 research and innovation program under grant agreement No. 820423 (S2QUIP), the European Research Council (ERC) under the European Union’s Horizon 2020 Research and Innovation Programme (SPQRel, grant agreement no. 679183), the Austrian Science Fund (FWF): P 29603, P 30459, I 4320, the Linz Institute of Technology (LIT) and the LIT Lab for secure and correct systems, supported by the State of Upper Austria, the German Federal Ministry of Education and Research via the funding program Photonics Research Germany (contract number 13N14846), Q.Com (Project No. 16KIS0110) and Q.Link.X (Project No. 16KIS0874), the DFG via SQAM (Project No. F1 947/5-1), the Nanosystem Initiative Munich, the MCQST, the Knut and Alice Wallenberg Foundation grant ”Quantum  Sensors”, the Swedish Research Council (VR) through the VR grant for international recruitment of leading researchers (Ref: 2013-7152), and Linn\ae{}us Excellence Center ADOPT.
K.M. acknowledges support from the Bavarian Academy of Sciences and Humanities. K.D.J. acknowledges funding from the Swedish Research Council (VR) via the starting grant HyQRep (Ref.: 2018-04812). J.V. acknowledges support from the National Science Foundation, Award number ECCS 1839056. K. D. J. acknowledges fruitful discussions with Stefan Schumacher.
A.R. acknowledges fruitful discussions with Y. Huo, G. Weihs, R. Keil and S. Portalupi. 
\end{acknowledgments}

%\bibliographystyle{naturemag}
%
%\bibliography{TPEvsRF}

\end{document}